\documentclass[pdflatex,sn-mathphys]{sn-jnl}

\jyear{2021}%

\theoremstyle{thmstyleone}%
\theoremstyle{thmstyletwo}%

\theoremstyle{thmstylethree}%
\usepackage{comment}
\begin{document}

\title[]{Galaxies in voids assemble their stars slowly}


\author*[1]{\fnm{Jes\'us} \sur{Dom\'inguez-G\'omez}}\email{jesusdg@ugr.es}
\author[1,2]{\fnm{Isabel} \sur{P\'erez}}
\author[1,3]{\fnm{Tom\'as} \sur{Ruiz-Lara}}
\author[3]{\fnm{Reynier F.} \sur{Peletier}}
\author[4]{\fnm{Patricia} \sur{S\'anchez-Bl\'azquez}}
\author[1,2]{\fnm{Ute} \sur{Lisenfeld}}
\author[5,6]{\fnm{Jes\'us} \sur{Falc\'on-Barroso}}
\author[1]{\fnm{Manuel} \sur{Alc\'azar-Laynez}}
\author[1,2]{\fnm{Mar\'ia} \sur{Argudo-Fern\'andez}}
\author[7]{\fnm{Guillermo} \sur{Bl\'azquez-Calero}}
\author[8]{\fnm{H\'el\`ene} \sur{Courtois}}
\author[1,9,7]{\fnm{Salvador} \sur{Duarte Puertas}}
\author[1,2]{\fnm{Daniel} \sur{Espada}}
\author[1,2]{\fnm{Estrella} \sur{Florido}}
\author[7]{\fnm{Rub\'en} \sur{Garc\'ia-Benito}}
\author[1]{\fnm{Andoni} \sur{Jim\'enez}}
\author[10]{\fnm{Kathryn} \sur{Kreckel}}
\author[1,2]{\fnm{M\'onica} \sur{Relaño}}
\author[1,2]{\fnm{Laura} \sur{S\'anchez-Menguiano}}
\author[3]{\fnm{Thijs} \spfx{van der} \sur{Hulst}}
\author[3]{\fnm{Rien} \spfx{van de} \sur{Weygaert}}
\author[1,2]{\fnm{Simon} \sur{Verley}}
\author[1,2]{\fnm{Almudena} \sur{Zurita}}

\affil*[1]{\orgdiv{Departamento de F\'isica Te\'orica y del Cosmos}, \orgname{Universidad de Granada}, \orgaddress{\street{Campus Fuente Nueva, Edificio Mecenas}, \city{Granada}, \postcode{E-18071}, \country{Spain}}}

\affil[2]{\orgdiv{Instituto Carlos I de F\'isica Te\'orica y Computacional}, \orgname{Universidad de Granada}, \orgaddress{\street{Campus Fuente Nueva, Facultad de Ciencias}, \city{Granada}, \postcode{E-18071}, \country{Spain}}}

\affil[3]{\orgdiv{Kapteyn Astronomical Institute}, \orgname{University of Groningen}, \orgaddress{\street{Landleven 12}, \city{AD Groningen}, \postcode{9747}, \country{The Netherlands}}}


\affil[4]{\orgdiv{Departamento de F\'isica de la Tierra y Astrof\'isica \& IPARCOS}, \orgname{Universidad Complutense de Madrid}, \orgaddress{\city{Madrid}, \postcode{E-28040}, \country{Spain}}}

\affil[5]{\orgname{Instituto de Astrofísica de Canarias}, \orgaddress{\street{Vía Láctea s/n}, \city{La Laguna, Tenerife}, \postcode{38205}, \country{Spain}}}

\affil[6]{\orgdiv{Departamento de Astrofísica}, \orgname{Universidad de La Laguna}, \orgaddress{\street{Vía Láctea s/n}, \city{La Laguna, Tenerife}, \postcode{38200}, \country{Spain}}}

\affil[7]{\orgname{Instituto de Astrofísica de Andalucía - CSIC}, \orgaddress{\street{Glorieta de la Astronomía s.n.}, \city{Granada}, \postcode{18008}, \country{Spain}}}

\affil[8]{\orgname{Universit\'e Claude Bernard Lyon 1, IUF, IP2I Lyon}, \orgaddress{\street{4 rue Enrico Fermi}, \city{Villeurbanne}, \postcode{69622}, \country{France}}}

\affil[9]{\orgdiv{D\'epartement de Physique, de G\'enie Physique et d’Optique}, \orgname{Universit\'e Laval, and Centre de Recherche en Astrophysique du Qu\'ebec (CRAQ)}, \orgaddress{\city{Qu\'ebec, QC}, \postcode{G1V 0A6}, \country{Canada}}}

\affil[10]{\orgdiv{Astronomisches Rechen-Institut, Zentrum für Astronomie}, \orgname{Universität Heidelberg}, \orgaddress{\street{Mönchhofstraße 12-14}, \city{Heidelberg}, \postcode{ D-69120}, \country{Germany}}}


\abstract{\bf Galaxies in the Universe are distributed in a web-like structure characterised by different large-scale environments: dense clusters, elongated filaments, sheetlike walls, and under-dense regions, called voids\cite{2001ApJ...557..495P, 2011AJ....141....4K, 2012MNRAS.421..926P, 2012ApJ...744...82V, 2016IAUS..308..493V}.
The low density in voids is expected to affect the properties of their galaxies. Indeed, previous studies\cite{2004ApJ...617...50R, 2005ApJ...624..571R,  2006MNRAS.372.1710P, 2007ApJ...658..898P, 2012MNRAS.426.3041H, 2012AJ....144...16K, 2014MNRAS.445.4045R, 2016MNRAS.458..394B, 2021ApJ...906...97F} have shown that galaxies in voids are on average bluer and less  massive, and have later morphologies and higher current star formation rates than galaxies in denser large-scale environments.
However, it has never been observationally proved that the star formation histories (SFHs) in void galaxies are substantially different from those in filaments, walls, and clusters. 
Here we show that void galaxies have had, on average, slower SFHs than galaxies in denser large-scale environments.
We also find two main SFH types present in all the environments:  'short-timescale' galaxies are not affected by their large-scale environment at early times but only later in their lives; 'long-timescale' galaxies have been continuously affected by their environment and stellar mass. Both types have evolved slower in voids than in filaments, walls, and clusters.
}




\maketitle

\begin{figure}
\centering
\includegraphics[width=\linewidth]{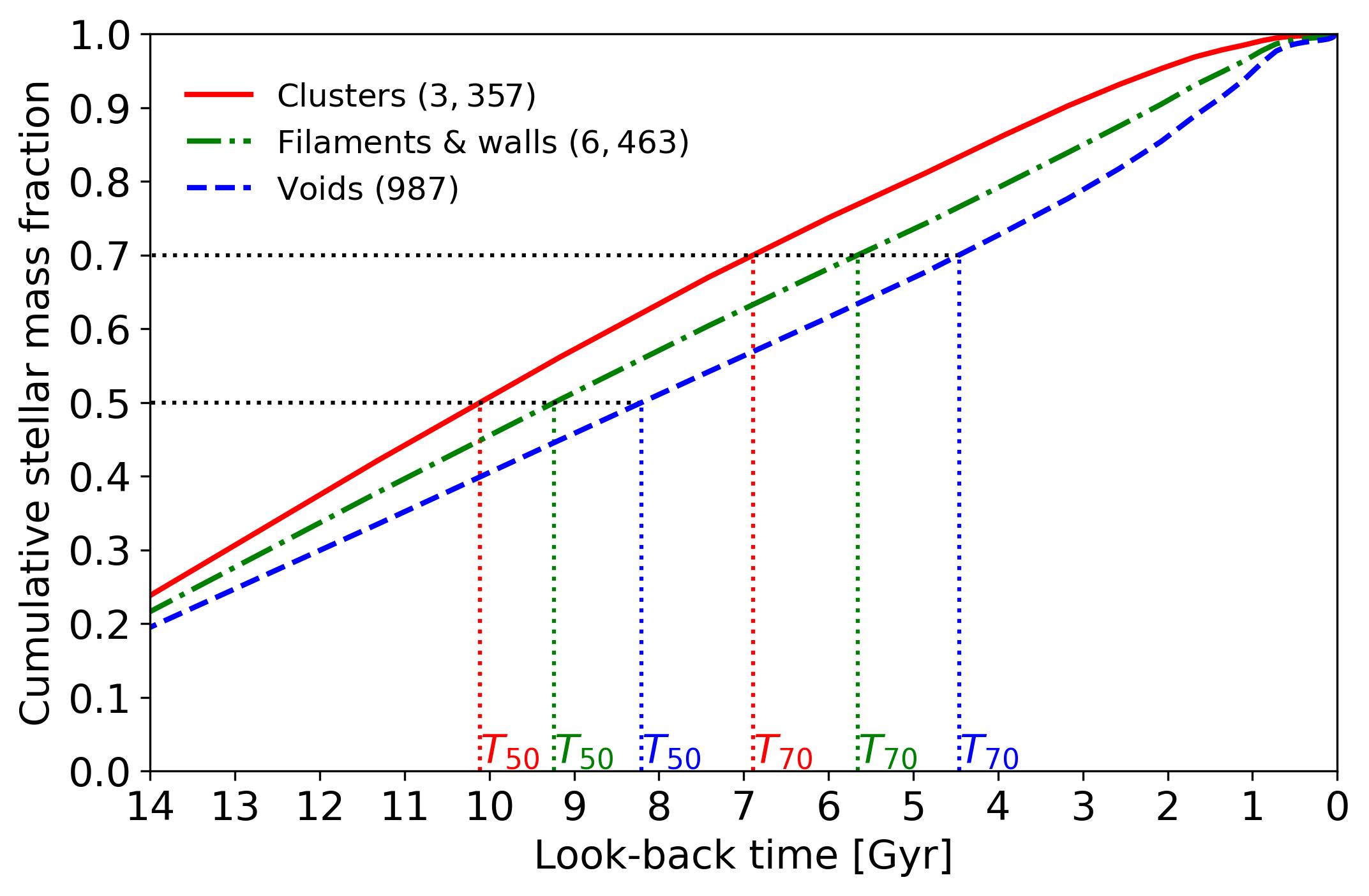} 
\caption{{\bf Average cumulative star formation history.} Cumulative stellar mass fraction formed at a given look-back time (in Giga years, Gyr), for galaxies in voids (blue dashed line), filaments \& walls (green dot-dashed line), and clusters (red solid line). The used samples take into account the selection criteria applied by the quality control analysis and are based on different stellar mass distributions. As representative quantities of the stellar mass assembly rate, dotted lines depict the average assembly times of 50\% ($T_{\rm 50}$) and 70\% ($T_{\rm 70}$) of the stellar mass, for which we find that the difference in SFH are maximal. By definition, $T_{\rm 100}=0\;{\rm Gyr}$ means that the galaxy forms 100\% of its stellar mass today, and $T_{\rm 70}$ and $T_{\rm 50}$ are correlated, i.e. if $T_{\rm 50}$ is higher in a sample, the $T_{\rm 70}$ is also higher. In general, the standard error of the mean (s.e.m., $1\sigma$) is smaller than the line width of the curves. The number of galaxies are represented in the legend for each large-scale environment.}
\label{fig:cum_sfh}
\end{figure}
\begin{figure}
\centering
\includegraphics[width=\linewidth]{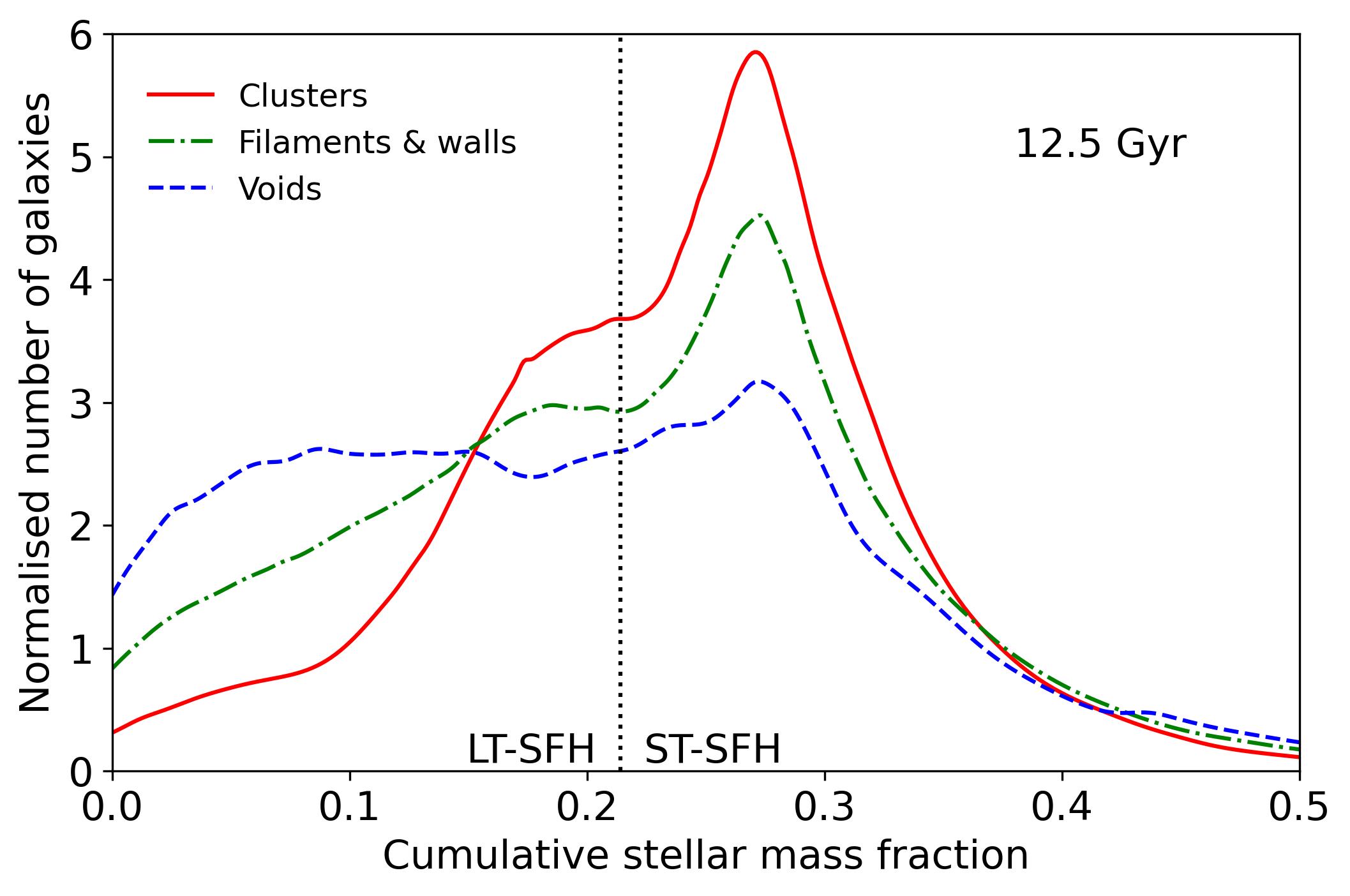} 
\caption{{\bf Bimodal distributions of the cumulative star formation histories at 12.5 Gyr.} Normalised number of galaxies vs. the cumulative stellar mass fraction. The used samples take into account the selection criteria applied by the quality control analysis and are based on different stellar mass distributions. The SFHs are classified into ST-SFH or LT-SFH using the vertical dotted line at 21.4\% (relative minimum of the sample of galaxies in filament \& walls and the inflexion point of the sample of galaxies in clusters) of the stellar mass as a classification criterion, which splits the distributions into two areas, which represent the probability of a galaxy to have one of the SFH types in each large-scale environment.}
\label{fig:SFHtype}
\end{figure}

In Figure \ref{fig:cum_sfh} we find that, on average, galaxies in voids assemble 50\% and 70\% of their stellar mass later than in filaments \& walls by $1.03\pm0.06$ and $1.20\pm0.05\,{\rm Gyr}$, respectively, and much later than in clusters (by $1.91\pm0.06$  and $2.43\pm0.05\,{\rm Gyr}$). 
In addition, in Figure \ref{fig:SFHtype} we find that the SFHs at early times describe a bimodal distribution around this average in the three large-scale environments.
We then classify the SFHs in two types: the short-timescale SFH (ST-SFH) is characterised by a high star formation ($\sim$27\% of the total stellar mass, the peak of the distributions) happening at the earliest time, while the long-timescale SFH  (LT-SFH) has a star formation happening more uniformly over time.
The distinction between these two types of SFHs allows us to evaluate the role of the large-scale environment in star formation, comparing the shape of the SFHs and assembly times between all three large-scale environments, and also paying attention to the probability of finding each SFH type in voids, filaments \& walls, and clusters (see Extended Data Figure \ref{fig:sfh_example} for the exact shape of an example of an ST-SFH galaxy and an LT-SFH galaxy and the differences between them). 

\begin{figure}
\centering
\includegraphics[width=\linewidth]{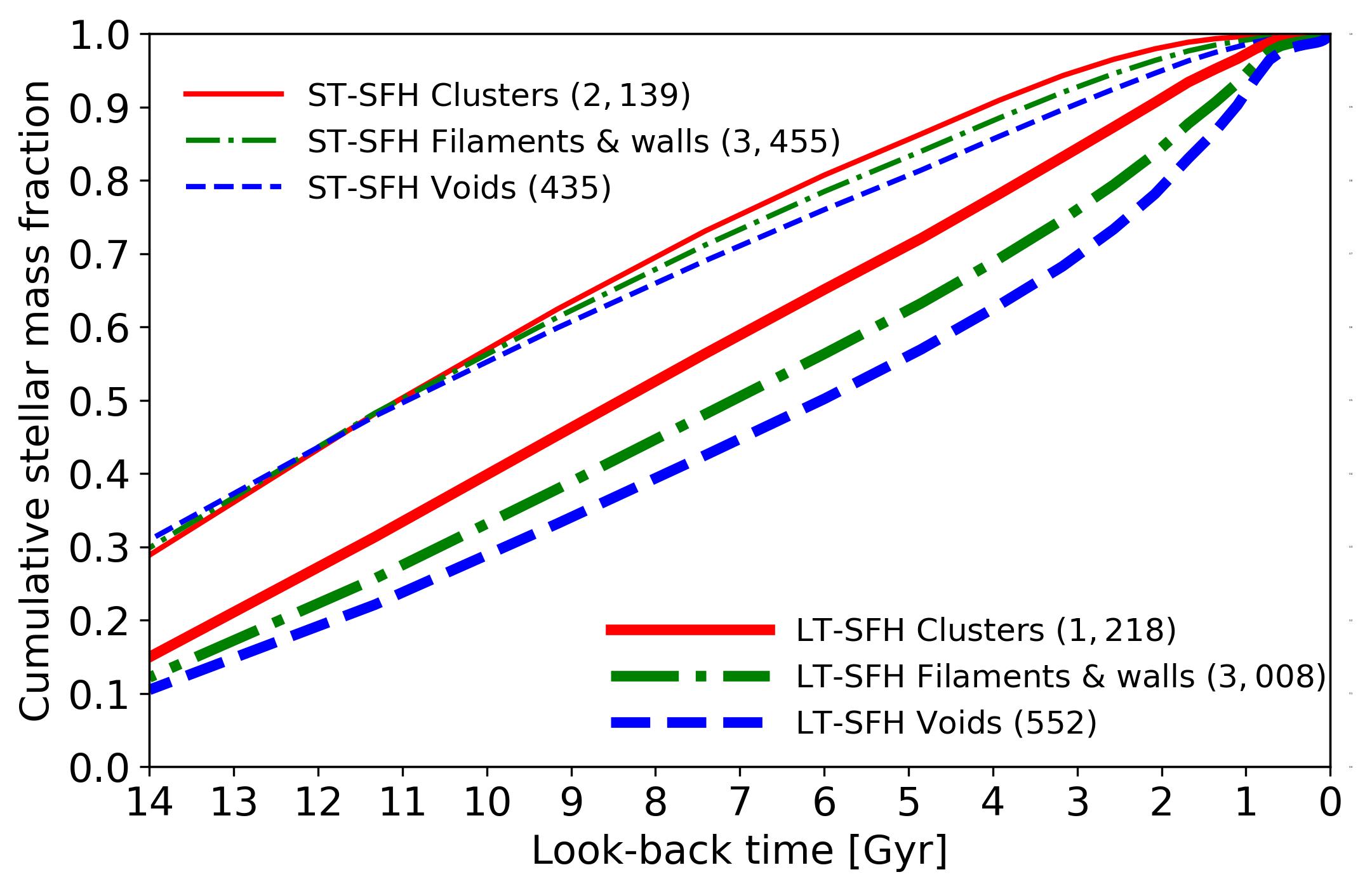 }
\caption{{\bf Tow types of star formation histories.} Cumulative stellar mass fraction formed at a given look-back time (in Giga years, Gyr) for galaxies with ST-SFH (thin lines) and LT-SFH (thick lines), in voids (blue dashed lines), filaments \& walls (green dot-dashed lines), and clusters (red solid lines). The used samples take into account the selection criteria applied by the quality control analysis and are based on different stellar mass distributions. In general, the s.e.m. ($1\sigma$) is smaller than the line width of the curves. The number of galaxies for each large-scale environment and SFH type are given in the legends.}
\label{fig:cum_sfh_type}
\end{figure}

We see in Figure \ref{fig:cum_sfh_type} that galaxies with ST-SFHs assembled on average 30\% of their stellar mass at early times \makebox{($\sim$ 12.5 Gyr ago)}, and decrease their star formation later in their lives. Galaxies with LT-SFHs, however, have assembled a lower stellar mass fraction ($\sim$15\%) at early times. ST-SFH galaxies, by definition, assemble their stellar mass earlier than LT-SFH. It is more likely for a void galaxy to have a LT-SFH ($51.7\pm0.9\%$) than for those in filaments \& walls ($44.5\pm0.3\%$) or clusters ($36.1\pm0.5\%$, see legends in Figure \ref{fig:cum_sfh_type}). Galaxies with ST-SFHs, on average, assemble their stellar mass at similar rates in the three large-scale environments. Galaxies with LT-SFHs, on average, assemble 50\% and 70\% of their stellar mass slower in voids than in filaments \& walls by $1.05\pm0.09$ and $0.86\pm0.06\,{\rm Gyr}$, respectively, and much slower than in clusters (by $2.38\pm0.10$ and $2.22\pm0.07\,{\rm Gyr}$). These might be the main reasons why we find in Figure \ref{fig:cum_sfh} that galaxies, on average, assemble their stellar mass later in voids than in denser large-scale environments. However, the stellar mass distributions of these galaxy samples depend on the large-scale environment\cite{2005ApJ...624..571R} (galaxies in voids are on average less massive than galaxies in denser large-scale environments, see Extended Data Figures \ref{fig:col_mag_all} and \ref{fig:col_mag_mask}), and it is necessary to test how these differences affect our results carrying out further comparisons for a given stellar mass. 

\begin{figure*}
\centering
\includegraphics[width=\linewidth]{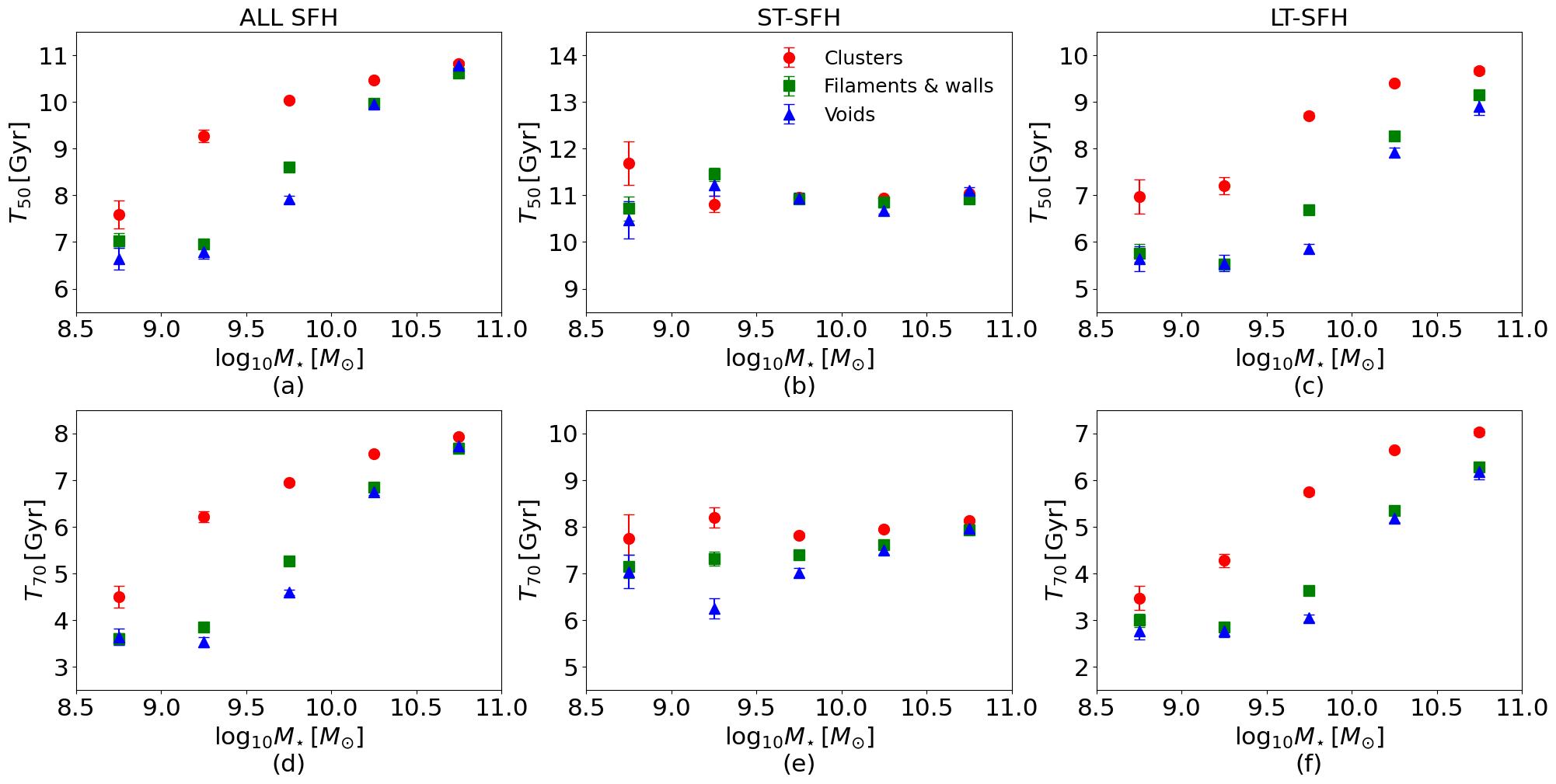} 
\caption{{\bf Median assembly times vs. stellar mass.} Assembly time (in Giga years, Gyr) of 50\% ($T_{\rm 50}$) and 70\% ($T_{\rm 70}$) of the stellar mass for all the SFH types (a and d, respectively), galaxies with ST-SFH (b and e), and LT-SFH (c, f), in voids (blue triangles), filaments \& walls (green squares), and clusters (red circles). The used samples take into account the selection criteria applied by the quality control analysis and are based on the same stellar mass distributions inside every stellar mass bin. The error bars represent the s.e.m. ($1\sigma$). The values and errors represented in this figure are shown in Extended Data Table \ref{tab:Tf_KS}, together with the number (and fraction of the SFH types) of galaxies in each
large-scale environment and stellar mass bin.}
\label{fig:T_KS}
\end{figure*}

We then define three new sub-samples with the same stellar mass distribution (see Methods section ‘Sample selection’ for more details). 
In Figure \ref{fig:T_KS} (a and d) we see that, regardless of their SFH type, cluster galaxies, on average, assemble their stellar mass faster than galaxies in voids, and filaments \& walls at any given stellar mass, except for very high stellar masses, where galaxies assemble their stellar mass at the same time in all the three large-scale environments. Low-mass and high-mass galaxies, on average, assemble 50\% and 70\% of their stellar mass at the same rate in voids and filaments \& walls. 
However, intermediate-mass galaxies assemble both 50\% and 70\% of their stellar mass later in voids than in filaments \& walls.

We find in Figure \ref{fig:T_KS} (b) that galaxies with ST-SFHs, on average, have formed 50\% of their stars very early ($\rm \sim11\,Gyr$ ago), independently of their large-scale environment and stellar mass. This suggests that, in the early Universe, the density contrasts between the upcoming large-scale environments were not strong enough to create a difference in the assembly rate between the galaxies that were forming at that time. The assembly time differences that we find for galaxies with ST-SFHs are only imprinted later on in their evolution (e.g. $T_{\rm 70}$, see Figure \ref{fig:T_KS} (e)), when the large-scale environment does play a role. Whereas galaxies in clusters with ST-SFH exhibit similar assembly times for any given stellar mass ($ T_{\rm 70}\sim 8 \;{\rm Gyr}$), void and filament \& wall galaxies, at some point, slow down their evolution compared to cluster galaxies, more substantially at low stellar masses than at high stellar masses. This might indicate that low-mass galaxies with ST-SFHs were formed at early times as their massive counterparts but have been later affected by the large-scale environmental density, slowing down their SFH. However, at very low stellar masses the number of galaxies (8, 24, and 9 galaxies) is less statistically significant than at higher stellar masses. High-mass galaxies with ST-SFHs have been less affected by their large-scale environment due to early mergers (even in voids, whose galaxies are not necessarily isolated, as they can be found in groups and in mergers\cite{1996AJ....111.2150S, 2012AJ....144...16K, 2013AJ....145..120B}), or the effect of more massive dark matter halos. Moreover, the fraction of ST-SFH low-mass galaxies (see Extended Data Table \ref{tab:Tf_KS}) is much lower than the fraction of ST-SFH high-mass galaxies, which suggests that galaxies that were assembled quickly at the very beginning of the Universe (ST-SFH) are more likely to be massive galaxies now, to have gathered mass by consecutive mergers, to have more massive dark matter halos, to run out of gas, and to quench.

Galaxies with LT-SFHs have assembled their stellar mass later than ST-SFH by $\rm 1.09$ to $\rm 5.88\,Gyr$ depending on the large-scale environment, assembly time, and stellar mass. This delay might have been enough for the LT-SFH to be affected by the large-scale environment since very early, in contrast with ST-SFH. We find in Figure \ref{fig:T_KS} (c and f) that void galaxies with LT-SFHs evolve slower than galaxies in clusters at any given stellar mass, and slightly slower than galaxies in filaments \& walls at intermediate stellar masses. 
The evolution of cluster galaxies with LT-SFH are accelerated, at any given stellar mass, by the higher density and higher probability of undergoing interactions in their large-scale environment compared to voids, and filament \& walls. In the same way, galaxies in filaments \& walls evolve faster than galaxies in voids, indicating that evolution at later times is influenced by the large-scale structure, more substantially at intermediate stellar masses than at high and low stellar masses. At high stellar masses galaxies might have been more affected by local interactions or their massive dark matter halos than by the large-scale environments. 
Low-mass galaxies might have been captured as satellites of more massive galaxies, being more affected by local processes and by the central galaxies of their system, and less affected by their large-scale environment. Additionally, the lack of assembly time differences at low stellar masses may be understood within the Halo Occupation Distribution (HOD) paradigm.

We observe differences in the evolution of galaxies comparing their SFHs for different large-scale environments, SFH types, and stellar masses. Although the mechanics that generate these differences are not clear yet, we can identify several processes that might have triggered these SFH differences between galaxies in different large-scale environments. 
Differences in the halo-to-stellar mass ratio, the AGNs activity, and the gas accretion between the three large-scale environments might be some of the reasons why the SFHs in void galaxies are, on average, slower than in filaments \& walls, and much slower than in clusters (see Methods section ‘Additional discussion’ for a more detailed discussion).

\newcommand*\aap{Astron. Astrophys.}
\let\astap=\aap
\newcommand*\aapr{A\&A~Rev.}
\newcommand*\aaps{A\&AS}
\newcommand*\actaa{Acta Astron.}
\newcommand*\aj{Astron. J.}
\newcommand*\ao{Appl.~Opt.}
\let\applopt\ao
\newcommand*\apj{Astrophys. J.}
\newcommand*\apjl{Astrophys. J.}
\let\apjlett\apjl
\newcommand*\apjs{ApJS}
\let\apjsupp\apjs
\newcommand*\aplett{Astrophys.~Lett.}
\newcommand*\apspr{Astrophys.~Space~Phys.~Res.}
\newcommand*\apss{Ap\&SS}
\newcommand*\araa{ARA\&A}
\newcommand*\azh{AZh}
\newcommand*\baas{BAAS}
\newcommand*\bac{Bull. astr. Inst. Czechosl.}
\newcommand*\bain{Bull.~Astron.~Inst.~Netherlands}
\newcommand*\caa{Chinese Astron. Astrophys.}
\newcommand*\cjaa{Chinese J. Astron. Astrophys.}
\newcommand*\fcp{Fund.~Cosmic~Phys.}
\newcommand*\gca{Geochim.~Cosmochim.~Acta}
\newcommand*\grl{Geophys.~Res.~Lett.}
\newcommand*\iaucirc{IAU~Circ.}
\newcommand*\icarus{Icarus}
\newcommand*\jcap{J. Cosmology Astropart. Phys.}
\newcommand*\jcp{J.~Chem.~Phys.}
\newcommand*\jgr{J.~Geophys.~Res.}
\newcommand*\jqsrt{J.~Quant.~Spectr.~Rad.~Transf.}
\newcommand*\jrasc{JRASC}
\newcommand*\memras{MmRAS}
\newcommand*\memsai{Mem.~Soc.~Astron.~Italiana}
\newcommand*\mnras{Mon. Not. R. Astron. Soc.}
\newcommand*\na{New A}
\newcommand*\nar{New A Rev.}
\newcommand*\nat{Nature}
\newcommand*\nphysa{Nucl.~Phys.~A}
\newcommand*\pasa{PASA}
\newcommand*\pasj{PASJ}
\newcommand*\pasp{PASP}
\newcommand*\physrep{Phys.~Rep.}
\newcommand*\physscr{Phys.~Scr}
\newcommand*\planss{Planet.~Space~Sci.}
\newcommand*\pra{Phys.~Rev.~A}
\newcommand*\prb{Phys.~Rev.~B}
\newcommand*\prc{Phys.~Rev.~C}
\newcommand*\prd{Phys.~Rev.~D}
\newcommand*\pre{Phys.~Rev.~E}
\newcommand*\prl{Phys.~Rev.~Lett.}
\newcommand*\procspie{Proc.~SPIE}
\newcommand*\qjras{QJRAS}
\newcommand*\rmxaa{Rev. Mexicana Astron. Astrofis.}
\newcommand*\skytel{S\&T}
\newcommand*\solphys{Sol.~Phys.}
\newcommand*\sovast{Soviet~Ast.}
\newcommand*\ssr{Space~Sci.~Rev.}
\newcommand*\zap{ZAp}


\clearpage

\clearpage
\subsection*{Acknowledgements} 
We are grateful to the referees that have helped us to improved the readability and clarity of the text.
We acknowledge financial support by the research projects AYA2017-84897-P, PID2020-113689GB-I00, and PID2020-114414GB-I00, financed by MCIN/AEI/10.13039/501100011033, the project A-FQM-510-UGR20 financed from FEDER/Junta de Andalucía-Consejer\'ia de Transforamción Económica, Industria, Conocimiento y Universidades/Proyecto and by the grants P20\_00334 and FQM108, financed by the Junta de Andaluc\'ia (Spain).
    TRL acknowledges support from Juan de la Cierva fellowship (IJC2020-043742-I), financed by MCIN/AEI/10.13039/501100011033.
    HC acknowledges support from the Institut Universitaire de France and the CNES. This work was supported by the Arqus European University and ANR program France 2030.
    LSM acknowledges support from Juan de la Cierva fellowship (IJC2019-041527-I) financed by MCIN/AEI/10.13039/501100011033.
    JFB  acknowledges support through the RAVET project by the grant PID2019-107427GB-C32 from the Spanish Ministry of Science, Innovation and Universities (MCIU), and through the IAC project TRACES which is partially supported through the state budget and the regional budget of the Consejer\'ia de Econom\'ia, Industria, Comercio y Conocimiento of the Canary Islands Autonomous Community.
    D.E. acknowledges support from a Beatriz Galindo senior fellowship (BG20/00224) from the Spanish Ministry of Science and Innovation.
    MAF acknowledges support the Emergia program (EMERGIA20\_38888) from Consejería de Transformación Económica, Industria, Conocimiento y Universidades and University of Granada.
    RGB acknowledges financial support from the grants CEX2021-001131-S funded by MCIN/AEI/10.13039/501100011033, SEV-2017-0709, to PID2019-109067-GB100, and grant 202250I003 "AYUDAS DE INCORPORACIÓN A CIENTÍFICOS TITULARES". 
    SDP acknowledges financial support from Juan de la Cierva Formaci\'on fellowship (FJC2021-047523-I) financed by MCIN/AEI/10.13039/501100011033 and by the European Union "NextGenerationEU"/PRTR, Ministerio de Econom\'ia y Competitividad under grant PID2019-107408GB-C44, from Junta de Andaluc\'ia Excellence Project P18-FR-2664, and also from the State Agency for Research of the Spanish MCIU through the `Center of Excellence Severo Ochoa' award for the Instituto de Astrof\'isica de Andaluc\'ia (SEV-2017-0709).
    G.B-C acknowledges financial support from grants PID2020-114461GB-I00 and CEX2021-001131-S, funded by MCIN/AEI/10.13039/501100011033, from Junta de Andalucía (Spain) grant P20-00880 (FEDER, EU) and from grant PRE2018-086111 funded by MCIN/AEI/10.13039/501100011033 and by 'ESF Investing in your future'.
    KK gratefully acknowledges funding from the Deutsche Forschungsgemeinschaft (DFG, German Research Foundation) in the form of an Emmy Noether Research Group (grant number KR4598/2-1, PI Kreckel).
    This research made use of Astropy, a community-developed core Python (http://www.python.org) package for Astronomy
    ; ipython
    ; matplotlib
    ; SciPy, a collection of open source software for scientific computing in Python
    ; APLpy, an open-source plotting package for Python
    ; and NumPy, a structure for efficient numerical computation
    .

\subsection*{Author contributions} 

\noindent JDG: corresponding author, sample selection, control sample selection, data analysis, results interpretation, and writing.
IP: sample selection, control sample selection, data analysis, results interpretation, and writing.
TRL: data analysis, results interpretation, and writing.
RFP: data analysis, results interpretation, and writing.
PSB: data analysis, results interpretation, and writing.
UL: sample selection, control sample selection, data analysis, results interpretation, and writing.
JFB: data analysis, results interpretation, and writing.
MAL: sample selection, results interpretation, and writing.
MAF: sample selection, results interpretation, and writing.
GBC: sample selection, results interpretation, and writing.
HC: results interpretation, and writing.
SDP: sample selection, control sample selection, results interpretation, and writing.
DE: results interpretation, and writing.
EF: sample selection, results interpretation, and writing.
RGB: results interpretation, and writing.
AJ: results interpretation, and writing.
KK: results interpretation, and writing.
MR: sample selection, results interpretation, and writing.
LSM: results interpretation, and writing.
TH: results interpretation, and writing.
RW: results interpretation, and writing.
SV: sample selection, control sample selection, results interpretation, and writing.
AZ: sample selection, results interpretation, and writing.

\clearpage

\section*{Methods}

\subsection*{Sample selection}

\noindent Our void galaxy sample is extracted from a nearby void galaxy catalogue defined by the Calar Alto Void Integral-field Treasury surveY (CAVITY \cite{CAVITYweb}) project.
This Calar Alto Observatory\cite{CAHAweb} legacy project is generating the first statistically complete IFU data set of void galaxies. The CAVITY survey presents a well defined selection of void galaxies, which are fully enclosed in the SDSS footprint within a 0.01 and 0.05 redshift range, and cover a wide range of stellar masses ($10^{8.5}\leq M_\star[{\rm M _\odot}]<10^{11.0}$).
The CAVITY survey aims to observe around $\sim300$ galaxies with the PMAS-PPAK integral field unit (IFU) at the Calar Alto Observatory, along with ancillary deep optical imaging, CO and HI data (Pérez et al. in prep.) to characterise the spatially resolved stellar populations and ionized gas content of void galaxies, together with their kinematics and dark mass assembly.
The CAVITY mother sample, which is made of 2,529 galaxies distributed in 15 voids, is a sub-sample of a previously defined catalogue of void galaxies presented in ref.\cite{2012MNRAS.421..926P}, where they apply the VoidFinder\cite{1997ApJ...491..421E, 2002ApJ...566..641H} algorithm to the galaxy distribution of the Sloan Digital Sky Survey (SDSS\cite{2009ApJS..182..543A}), and identify 79,947 galaxies in 1,055 voids with mean density contrast $\delta\rho/\rho=-0.94\pm0.03$ and radii larger than $10\,{\rm h^{-1}\,Mpc}$. The VoidFinder algorithm classifies as potential void galaxies those with their third nearest neighbour distance $d_3>6.3\,{\rm h^{-1}\,Mpc}$ and remove them from the SDSS galaxy sample, leaving only galaxies in denser large-scale environments. Then, it generates a space grid of cubic cells of size $5\,{\rm h^{-1}\,Mpc}$, identifies the empty ones as potential voids, and fits maximal spheres inside these empty regions. Spheres overlapping more than 10\% are unified in the same void, and galaxies inside these spheres are classified as void galaxies. The spheres with the same volume of the void defines its effective radius. However, voids are not perfectly spherical and some void galaxies lay beyond the effective radius of its void. There are many algorithms to find voids and other large-scale structures such as filaments, walls and clusters. These algorithms differ in the classification of the galaxies into different large-scale structures, specially when they are very close to the limit of the void, and some galaxies may have been potentially mis-classified as void galaxies. The CAVITY project carries out a selection of galaxies from this catalogue\cite{2012MNRAS.421..926P} that lay in the inner region of the voids (distance from the centre below 80\% of the effective radius of the void) to avoid the possible inclusion of galaxies inhabiting denser environments. Galaxies that satisfy the completeness (at least 20 galaxies distributed along the radius of the void) and observational (15 voids distributed along right ascension for continuous visibility along the year) requirements of the project form part of the CAVITY mother galaxy sample. 

The control sample comprises galaxies inhabiting large-scale environments in the nearby Universe that are denser than voids, these are filaments, walls, and clusters.
We divide it into two different samples one is made of galaxies in clusters, the other one is made of galaxies in filaments and walls. The number density of galaxies is very similar in filaments and walls, therefore, we consider both environments together (filaments \& walls hereafter), as the regions where galaxies are neither in voids nor in clusters.
We begin the selection with 109,945 objects that are classified in SDSS as galaxies, have available spectral data in the SDSS, and lie in the same redshift range as the CAVITY mother galaxy sample (0.01-0.05). We then remove 38,473 void galaxies by excluding objects that belong to the previously cited catalogue of void galaxies\cite{2012MNRAS.421..926P}. We then cross-match the remaining galaxies with a previously defined catalogue of groups of galaxies\cite{2017A&A...602A.100T} from SDSS (considering objects with 30 or more companions as cluster galaxies\cite{1989ApJS...70....1A}) and separate 6,189 objects, which make the cluster galaxy control sample. The remaining 65,283 galaxies are considered filament \& wall galaxies. Furthermore, to save computational time, we select from this sample a random sub-sample of 15,000 filament \& wall galaxies preserving the same $g-r$ colour, stellar mass, and redshift distributions (two-samp KS-test p-value $>0.95$). These three samples are magnitude-limited due to the SDSS completeness limit at r-Petrosian $<$17.77 mag\cite{2002AJ....124.1810S, 2015A&A...578A.110A}. This means that the sample is progressively less sensitive to faint objects with increasing redshift. However, given the type of study that we carry out and given that the narrow redshift range of the CAVITY sample, we do not expect this to be a severe problem for this work.

In Extended Data Fig. \ref{fig:col_mag_all} we compare colour and stellar mass distributions between the three samples. We see that part of the void galaxies are located in the red sequence but they mainly populate the blue cloud. This distribution is more balanced in filaments \& walls where galaxies are equally distributed between the red sequence and the blue cloud. However, the majority of cluster galaxies cover the red sequence, and only a small part of them are distributed over the blue cloud. As it has been observed before\cite{2004ApJ...617...50R, 2005ApJ...624..571R,  2006MNRAS.372.1710P, 2007ApJ...658..898P, 2012MNRAS.426.3041H, 2012AJ....144...16K, 2014MNRAS.445.4045R, 2016MNRAS.458..394B, 2021ApJ...906...97F} void galaxies are on average bluer and less massive than galaxies in filaments, walls, and clusters. 
Then, we proceed to carry out our analysis and quality control cuts (see Methods section ‘Quality control’ for more details) to remove some galaxies with bad-fitted spectra (mainly low mass objects from the blue cloud with low signal-to-noise) that modify the distributions (see Extended Data Figure \ref{fig:col_mag_mask}). We lose 1,542 (61\%) galaxies from voids; 8,537 (57\%) from filaments \& walls; and 2,832 (46\%) from clusters. The mean stellar mass of the removed galaxies are similar in the three environments ($10^{9.2\pm0.1}\;{\rm M_\odot}$ in voids, $10^{9.3\pm0.1}\;{\rm M_\odot}$ in filaments \& walls and $10^{9.5\pm0.1}\;{\rm M_\odot}$ in clusters). Finally, after this quality control we are left with 987 galaxies in voids, 6,463 in filaments \& walls, and 3,357 in clusters for our study. In Extended Data Figure \ref{fig:col_mag_mask} we compare the colour vs. stellar mass distribution of the samples after our quality control. These are the samples we use for the global SFH comparison in figures \ref{fig:cum_sfh}, \ref{fig:SFHtype}, and \ref{fig:cum_sfh_type}.

As we see in Extended Data Figures \ref{fig:col_mag_all} and \ref{fig:col_mag_mask}, the stellar mass distribution of these samples is not the same even after the quality control, void galaxies are on average less massive than galaxies in denser environments. Therefore, in order to extend the comparison for a given stellar mass, we define five stellar mass bins (with a 0.5 dex width) between $10^{8.5}$ and $10^{11.0}\,{\rm M_\odot}$, and
we generate two control sub-samples with the same stellar mass distribution as our void galaxy sample inside every stellar mass bin, and apply the two-sample Kolmogorov–Smirnov test (two-samp KS-test) with p-values above 0.95, to ensure the accuracy with which the stellar mass distributions are matched.
We have not been able to extend the bins at lower or higher stellar masses because the number of galaxies beyond these limits was not enough, in at least one of the environments, to define sub-samples with similar stellar mass distributions applying KS-test. We are left with 978 void galaxies (we lose 9 void galaxies that lay outside the stellar mass bins), 4,800 filament \& wall galaxies, and 2,570 cluster galaxies. These are the samples that we use in Figure \ref{fig:T_KS} to compare the assembly times for a given stellar mass.  

\subsection*{Spectral analysis}

In this study we recover the stellar line-of-sight velocity distribution (LOSVD, paying special attention to the recession velocity, $\rm V$, and the velocity dispersion, $\rm \sigma$), the emission lines, the stellar populations, and the SFHs of galaxies by fitting stellar spectrum templates to the observed spectrum of the galaxies. For our analysis we use: the synthetic spectral energy distributions (SEDs) for single-age, single-metallicity stellar populations (SSPs) of the Extended Medium resolution INT (Isaac Newton Telescope) Library of Empirical Spectra (E-MILES\cite{2006MNRAS.371..703S, 2011A&A...532A..95F, 2015MNRAS.449.1177V, 2016MNRAS.463.3409V}) as stellar spectrum templates; the observed galaxy spectra already available in the SDSS-DR7\cite{2009ApJS..182..543A} as data of analysis; the Penalized Pixel-Fitting  (pPXF\cite{2004PASP..116..138C, 2017MNRAS.466..798C, Cappellari2022}) as software to recover the stellar LOSVD and the emission lines of the gas; and the STEllar Content and Kinematics from high resolution galactic spectra via Maximum A Posteriori  (STECKMAP\cite{2006MNRAS.365...46O, 2006MNRAS.365...74O}) as algorithm to recover the stellar populations and SFHs.
    
The E-MILES\cite{2006MNRAS.371..703S, 2011A&A...532A..95F, 2015MNRAS.449.1177V, 2016MNRAS.463.3409V} SSP models are generated using the BaSTI\cite{2004ApJ...612..168P} isochrones and Kroupa\cite{2001MNRAS.322..231K} universal initial mass function (IMF). We expect that a change in the IMF would shift\cite{2019A&A...621A.120G} our results (assembly times $T_{50}$ and
$T_{70}$). However, as we assume the same IMF for the three large-scale environments, it will affect our
SFHs equally regardless of the environment, and thus, relative differences between voids, filaments
\& walls and clusters should remain. These models cover a wavelength range from 1,680 to 50,000 $\rm \AA$ with linear wavelength sampling with a pixel separation of 1.00 $\rm \AA$ and a variable instrumental dispersion of FWHM between 2.51 and 23.57 $\rm \AA$, which is constant (2.51 $\rm \AA$) inside the fitting wavelength range ($\rm 3,750 - 5,450 \,\rm \AA$) of our analysis.
    
The SDSS-DR7\cite{2009ApJS..182..543A} contains integrated optical spectra (fibre aperture with 3 arcsec diameter) for 1.6 million objects, including 930,000 galaxies, 120,000 quasars, and 460,000 stars observed at the Apache Point Observatory (APO) 2.5 m telescope.
The SDSS spectra have a wavelength coverage from 3,800 to 9,200 $\rm \AA$, logarithmic wavelength sampling with a pixel separation of $\rm 69\,km\,s^{-1}$ ($\rm \Delta \log_{10}(\lambda)=10^{-4}\,dex$), and a variable spectral power resolution ranging from $\rm  R\sim1,500$ at 3,800 $\rm \AA$ to  $\rm  R\sim2,500$ at 9,000 $\rm \AA$. The instrumental dispersion of the SDSS spectra is variable \makebox{(FWHM $\rm \sim 2.00-3.00\, \AA $)} inside the fitting wavelength range ($\rm 3,750 - 5,450 \,\rm \AA$) of our analysis, and it is different for every galaxy. This is taken into account later in the analysis. 

The pPXF\cite{2004PASP..116..138C, 2017MNRAS.466..798C, Cappellari2022} algorithm implements a non-parametric full spectral fitting technique to recover the LOSVD of stars and emission lines. We assume Gaussian-Hermite LOSVD for the stars, and a pure Gaussian LOSVD for the emission lines. This algorithm uses stellar and gas spectral templates, chooses a combination of them, and convolves them with the LOSVD that better fit the spectrum of the galaxy. We use the E-MILES stellar templates and, for the emission lines, we place several lines in the same template with fixed relative fluxes of emission-line doublets or Balmer series. The E-MILES stellar templates are synthetic or have been observed with different instruments than the one used to observe the spectrum of the galaxy. This means that the wavelength sampling and instrumental dispersion of the stellar templates (linear sampling and $ \rm FWHM = 2.51 \AA$ of instrumental dispersion) are different compared to the spectrum of the galaxy (logarithmic and $\rm FWHM  \sim 2.00-3.00\, \AA $, respectively). We re-sample the templates and convolve both the templates and the observed spectra of the galaxies to have the same wavelength sampling (logarithmic) and instrumental dispersion (3.00 \AA) in all of them. 

The STECKMAP\cite{2006MNRAS.365...46O, 2006MNRAS.365...74O} algorithm recovers the stellar populations of a galaxy as a combination of SSPs that are fitted to the observed spectrum of the galaxies, after removing the emission lines and assuming fixed stellar LOSVD (both previously derived with pPXF). From this combination of SSPs we derive the stellar mass fraction, metallicity, and age of the currently living stars of the galaxy. Afterwards, we apply a correction factor (which depends on the age and metallicity of each stellar population and is provided by the MILES group\cite{MILESweb}) to the current stellar mass fractions, in order to take into account the stars that were formed at a given cosmic look-back time but are not alive any more. For this purpose we follow the prescriptions in ref.\cite{1996ApJS..106..307V, 2001MNRAS.320..193B, 2010MNRAS.404.1639V}, using BaSTI\cite{2004ApJ...612..168P} isochrones and Kroupa\cite{2001MNRAS.322..231K} IMF. Finally, from these stellar population ages and corrected stellar mass fractions, we derive the SFH of a galaxy, as the cumulative stellar mass fraction formed at a given look-back time, and estimate its errors as the standard deviation of 5 Monte Carlo solutions of STECKMAP. Additionally, we interpolate over the cumulative SFH to calculate the times when 50\% ($\rm T_{50}$) and 70\% ($\rm T_{70}$) of the stellar mass of the galaxy was formed. We repeat this for the 5 Monte Carlo solutions and estimate the errors as the standard deviation. Examples of SFHs are shown in Extended Data Figure \ref{fig:sfh_example} for two individual galaxies.

\subsection*{Quality control}

After applying this analysis to our samples of galaxies in voids (2,545 objects), filaments \& walls (15,000), and clusters (6,189), we carry out a quality control to identify and remove the bad-fitted spectra. There are two main aspects that affect the fit quality: the signal-to-noise ratio (S/N) of the spectra and the intensity of the emission lines. The fit residual, which is the difference between the observed and the fitted spectra, is a good indicator of the fit quality. A high fit residual means that the observed spectrum is noisy or the fitted spectrum does not match the real one. In Extended Data Figure \ref{fig:sn_res} (a) we show the standard deviation of the fit residual normalised by the level of continuum around $\rm H\beta$ (second emission line of the Balmer series of the hydrogen atom, $\rm\sigma_{res}(H\beta)/Cont$) vs. the S/N in the continuum, the equivalent width of $\rm H\beta$ ($\rm \Delta H\beta_{eq}$) is colour-coded. Here we see how the  fit quality (stellar and gas emission) is affected by the S/N. The level of fit residual, relative to the continuum, decreases with S/N. We considered that a S/N $\geq$ 20 provides a good quality fit with a level of residual lower than 2\% of the continuum for the great majority of the galaxies. However, there is a group of galaxies with really intense or wide emission lines that are not well fitted although they have high S/N.

The pPXF algorithm is not efficient fitting very intense or non-gaussian emission lines and may generate high fit residuals. These residuals are small compared to the continuum (below 2\%) in bright galaxies but may be higher than the level of noise, leaving wavy features in the spectrum that may affect the fit of stellar populations. In Extended Data Figure \ref{fig:sn_res} (b) we show residual-to-noise ratio as the standard deviation of the residual at $\rm H\beta$ normalised by standard deviation of noise in the continuum next to $\rm H\beta$ ($\rm \sigma_{res}(H\beta)/\sigma_{noise}(H\beta)$), vs. the S/N in the continuum, $\rm \Delta H\beta_{eq}$ is colour-coded. Here we see that some bright galaxies (with intense emission lines in general) have the level of residual much higher than the level of noise. After a visual inspection of a set of these galaxies, we find featured residuals due to asymmetric, wide, or not gaussian emission lines (see good and bad fit examples in Extended Data Figure \ref{fig:view_fit}), and consider that $\rm \sigma_{res}(H\beta)/\sigma_{noise}(H\beta)\leq2.0$ provides good quality fit. However, this is only a small fraction of the total number of galaxies (5\% in voids, 8\% in filaments \& walls, and 7\% in clusters).

Additionally, as final step in our quality control, we take into account the aperture effect in the SDSS spectra. The optical spectra are available for the central region of our galaxies (fibre aperture with 3 arcsec diameter). This might introduce a bias for samples covering a large redshift range where this aperture would cover only the inner region of the nearby objects, but a large fraction of the galaxy for the distant ones. However, the apparent radius distribution (see Extended Data Figure \ref{fig:hist_R90}) is similar for the three samples and the redshift range (0.01 to 0.05) is the same, with absolute apertures ranging from 0.3 to 1.6 $\rm kpc$. 
We remove from our samples galaxies with $\rm R_{90r}>20\, arcsec$, for which the aperture effect would have a larger influence, to minimise a possible size bias in our study. The aperture effect only affects our study in the sense that our results are only valid for the centre of the galaxy.

\subsection*{Additional discussion}

In Figure \ref{fig:SFHtype} we show that the SFH of a galaxy can be classified as ST-SFH or LT-SFH. Although it might be tempting to associate the SFH bi-modality with the current galaxy colour, morphology or local environment, the SFH types do not do not correspond to the bimodal colour or morphology distributions (see Extended Data Figure \ref{fig:colour_mstar_types}), and do not strongly correlates with the local density. The SFH type determine the evolution of a galaxy in general, along its entire life. However, the colour and morphology are associated with current properties of the galaxy, which should not be that strongly affected by the overall SFH type (shape) but by the most recent stages of the SFH or the current environmental state or physical conditions (gas content, interactions, accretion, etc.)

Additionally, in Figure \ref{fig:T_KS} (c and f) we do not find assembly time differences for low-mas galaxies. This observation can be interpreted based on the Halo Occupation Distribution (HOD) paradigm as mentioned in the main text. According to this paradigm, galaxies in voids have higher halo mass ($\sim$ 10 \%)\cite{2020A&A...638A..60A} than in denser environments for a given stellar mass. This, together with the lower probability of finding high-mass galaxies in voids, makes low-mass galaxies in voids more likely to be the central objects of a system. However, low-mass galaxies in voids, which would have presumably evolved slower than galaxies in filaments \& walls due to their large-scale environment, might have compensated for these SFH differences by accelerating their star formation due to their higher halo masses. 

Our main finding is that galaxies assemble their stellar mass in voids slower than in filaments \& walls, and much slower than in clusters (see Figure \ref{fig:T_KS}). Here we discuss some of the physical processes that might have triggered these differences. Previous cosmological simulation analysis\cite{2018MNRAS.480.3978A, 2020A&A...638A..60A, 2020MNRAS.493..899H, 2022MNRAS.tmp.2385R} find that the halo-to-stellar mass ratio is higher in void galaxies compared to galaxies in denser large-scale environments, suggesting that galaxies in voids evolve slower than in filaments, walls, and clusters. 
Regarding the effect of active galactic nuclei (AGNs), some observational studies find statistical evidences for a larger fraction of AGNs \cite{2008ApJ...673..715C, 2022MNRAS.509.1805C}, and massive black holes (BH) \cite{2022MNRAS.509.1805C} in voids. However, these differences in the fraction of AGNs between galaxies in voids and in denser environments are not present in other observational studies\cite{2019ApJ...874..140A}. No differences in the BH-to-galaxy mass ratio are reported either by other simulation analysis\cite{2020MNRAS.493..899H}.
There is no agreement on how the large-scale environment affects the nuclear activity of the galaxies, and it is controversial to consider  AGNs as a possible mechanism that triggered the SFH differences that we find. A simulation analysis\cite{2005MNRAS.363....2K} found that there are mainly two modes of gas accretion in galaxies. In the cold accretion mode, the gas flows along the filaments into the galaxy. It dominates in low density large-scale environments, low stellar mass galaxies, and at high redshifts. In the hot accretion mode, the virialised gas around the galaxy falls into it while it cools down. It dominates in cluster large-scale environments, massive galaxies, and at low redshifts. This suggests that the gas accretion has been different throughout the SFH between galaxies in voids and denser large-scale environments, and this might have introduced current gas content differences between them. Some other observational studies\cite{1996AJ....111.2150S, 2021ApJ...906...97F} do not find any atomic gas mass differences between voids and galaxies in denser large-scale environments but others\cite{2012AJ....144...16K, 2022A&A...658A.124D, 2022MNRAS.tmp.2385R} find a tentative lack of atomic gas in void galaxies at stellar masses above $\rm10^{9.5}\,M_{\odot}$, the same range where we find the LT-SFH differences. However, a lack of atomic gas in galaxies does not necessarily imply a lack of molecular gas\cite{1986ApJ...301L..13K, 1997AJ....114.1753S, 2015ApJ...815...40D, 2016MNRAS.459.3574C, 2016A&A...590A..27G, 2022A&A...658A.124D}, from which the stars are formed.

\subsection*{Data availability}
The analysis and results in this work are based on public data: SDSS query/CasJobs (\url{http://casjobs.sdss.org/casjobs/}), and SDSS spectra (\url{http://data.sdss.org/sas/dr16/sdss/spectro/redux/26/spectra/}). Interested researchers can reproduce our analysis following the steps in the Methods section and using the public data and codes. They can also compare with our results using the electronic spreadsheets associated with the figures in the main text. 

We do not place our results for individual galaxies into public repository at the moment because two PhD students inside the CAVITY project are using these results for their thesis. Additionally, a great effort has been required to carry out this analysis, and the CAVITY project plans to base many of their future works on these results. At a later stage, once the PhD projects are finished and convenient exploitation of the work within the collaboration is done, we plan to make an ample data set available for the community. We need to highlight the 'legacy' nature of this project, as agreed in the memorandum of understanding with the Calar Alto observatory, but first, we reserve our rights for an embargo period for the full exploitation of this project.

\subsection*{Code availability}
Codes that support the analysis in this study are publicly available: pPXF\cite{2004PASP..116..138C, 2017MNRAS.466..798C, Cappellari2022}  (\url{https://www-astro.physics.ox.ac.uk/~cappellari/software/}), STECKMAP\cite{2006MNRAS.365...46O, 2006MNRAS.365...74O} (\url{https://urldefense.com/v3/__https://github.com/pocvirk/STECKMAP__;!!D9dNQwwGXtA!VggZnNu4_e840FF17iVF0CW79nTSLkzJ53o14bQwryoS3l_alwG4PzL_OFaVMnHJ8UNWkXs5WYNJtvkFBU-3y7O2nofr$})

\subsection*{Author Information} 
The authors declare that they have no competing financial interests.
Correspondence and requests for materials should be addressed to JDG ({\it jesusdg@ugr.es}).


\clearpage
\section*{Extended Data}
\counterwithin{figure}{section}
\renewcommand{\thefigure}{\arabic{figure}}
\setcounter{figure}{0}
\renewcommand{\figurename}{Extended Data Figure}
\renewcommand{\tablename}{Extended Data Table}

\begin{figure}[!h]
    \centering    
        \includegraphics[width=\linewidth]{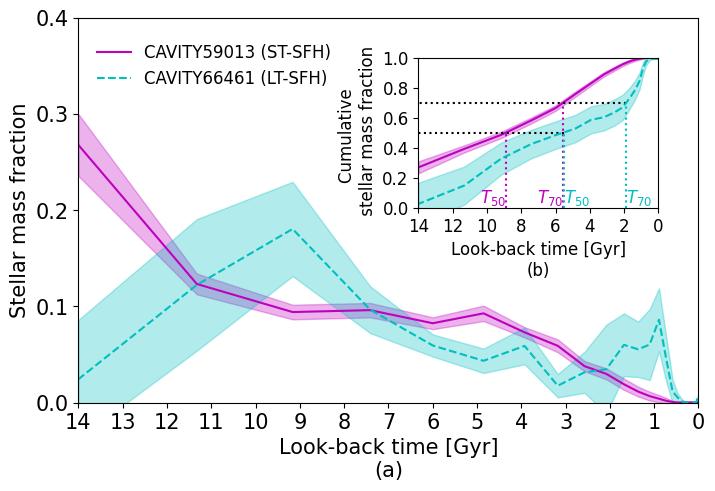}    
    \caption{{\bf Examples of star formation histories.} SFHs (a), and (b) cumulative SFHs for galaxies CAVITY59013 (solid magenta line) and  CAVITY66461 (dashed cyan line), which have ST-SFH and LT-SFH types, respectively. The shaded regions represent the errors of the stellar mass fraction of the SFH. The dotted lines in (b) represent the assembly times of the 50\% ($T_{\rm 50}$) and 70\% ($T_{\rm 70}$) of the stellar mass.}
    \label{fig:sfh_example}
\end{figure}

\begin{figure*}
    \centering
        \includegraphics[width=\linewidth]{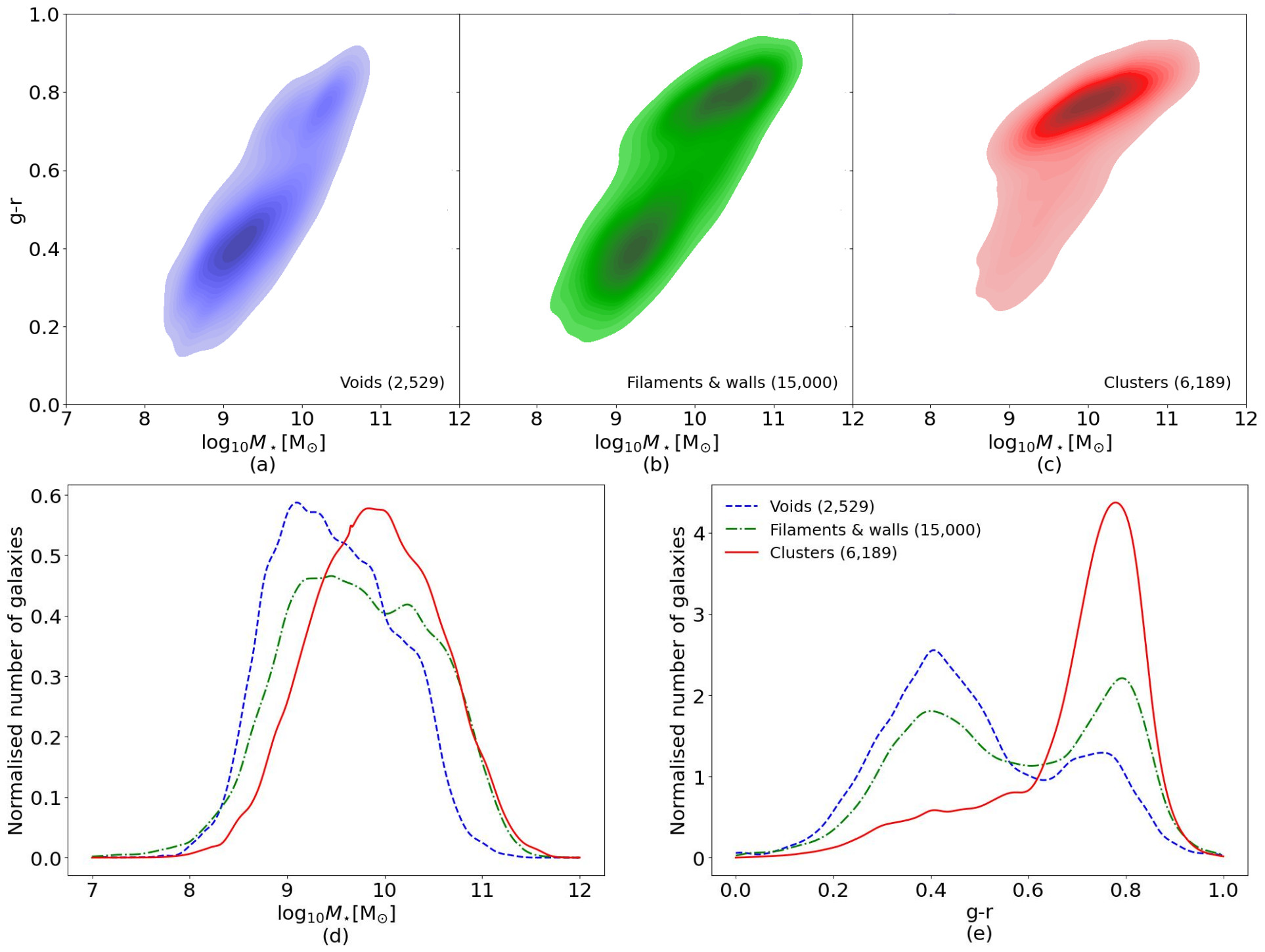} 
    \caption{{\bf Colour and stellar mass distribution before the quality control.} Color vs. stellar mass diagram for galaxies in voids (a), filaments (b), and clusters (c). Normalised distributions of the stellar mass (d) and $g-r$ colour (e) for galaxies in voids (blue dashed line), filaments \& walls (green dot-dashed line), and clusters (red solid line).}
    \label{fig:col_mag_all}
\end{figure*}

\begin{figure*}
    \centering
       \includegraphics[width=\linewidth]{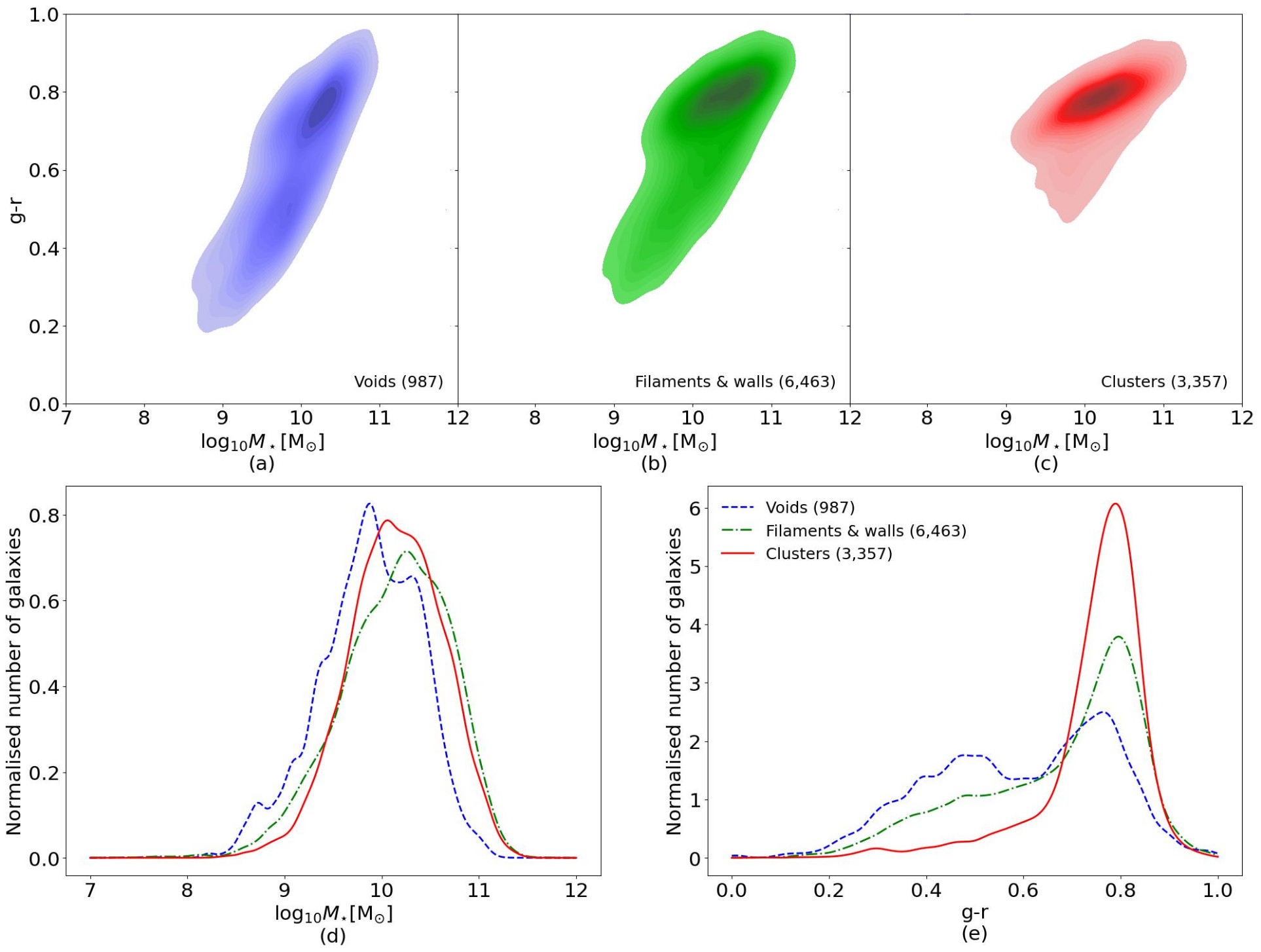}
    \caption{{\bf Colour and stellar mass distribution after the quality control.} Color vs. stellar mass diagram for galaxies in voids (a), filaments (b), and clusters (c). Normalised distributions of the stellar mass (d) and $g-r$ colour (e) for galaxies in voids (blue dashed line), filaments \& walls (green dot-dashed line), and clusters (red solid line).}
    \label{fig:col_mag_mask}
\end{figure*}

\begin{sidewaystable}
\sidewaystablefn%
\small
\begin{center}
\begin{minipage}{\textheight}
\begin{tabular}{c|ccc|ccc|ccc}
     & \multicolumn{3}{c|}{Voids}& \multicolumn{3}{c|}{Filaments \& walls}& \multicolumn{3}{c}{Clusters}\\
      $ \log_{10}M_\star{\rm [M_\odot]}$&  n(\%) & $ T_{\rm 50}\,{\rm[Gyr]}$ &   $ T_{\rm 70}\,{\rm[Gyr]}$ &  n(\%) & $ T_{\rm 50}\,{\rm[Gyr]}$ &   $ T_{\rm 70}\,{\rm[Gyr]}$&  n(\%) & $ T_{\rm 50}\,{\rm[Gyr]}$ &   $ T_{\rm 70}\,{\rm[Gyr]}$ \\
    \midrule
    \midrule
    \multicolumn{10}{l}{(a) All SFH}\\
    \midrule
       8.5-9.0 &        56 &   6.64$\pm$0.23 &  3.64$\pm$0.17 &          105 &   7.03$\pm$0.16 &   3.60$\pm$0.12 &          30 &   7.59$\pm$0.29 &  4.50$\pm$0.24 \\
       9.0-9.5 &       159 &   6.79$\pm$0.14 &  3.53$\pm$0.11 &          326 &   6.95$\pm$0.09 &   3.84$\pm$0.07 &         100 &   9.27$\pm$0.13 &  6.22$\pm$0.12 \\
      9.5-10.0 &       363 &   7.92$\pm$0.08 &  4.60$\pm$0.06 &         1500 &   8.61$\pm$0.04 &   5.26$\pm$0.03 &         855 &  10.03$\pm$0.04 &  6.94$\pm$0.04 \\
     10.0-10.5 &       309 &   9.95$\pm$0.06 &  6.74$\pm$0.05 &         2199 &   9.97$\pm$0.02 &   6.84$\pm$0.02 &        1212 &  10.47$\pm$0.02 &  7.57$\pm$0.02 \\
     10.5-11.0 &        91 &  10.79$\pm$0.07 &  7.73$\pm$0.07 &          670 &  10.62$\pm$0.03 &   7.69$\pm$0.03 &         373 &  10.82$\pm$0.03 &  7.93$\pm$0.03 \\
    
    \multicolumn{10}{l}{(b) ST-SFH}\\
    \midrule
       8.5-9.0 &    8(14\%) &  10.47$\pm$0.39 &  7.04$\pm$0.36 &      24(23\%) &  10.72$\pm$0.26 &   7.15$\pm$0.25 &      9(30\%) &  11.69$\pm$0.47 &  7.75$\pm$0.51 \\
   9.0-9.5 &   40(25\%) &  11.23$\pm$0.24 &  6.25$\pm$0.22 &      80(25\%) &  11.45$\pm$0.14 &   7.32$\pm$0.15 &     39(39\%) &  10.81$\pm$0.18 &  8.19$\pm$0.22 \\
  9.5-10.0 &  127(35\%) &  10.94$\pm$0.11 &  7.02$\pm$0.11 &     602(40\%) &  10.92$\pm$0.05 &   7.39$\pm$0.05 &    463(54\%) &  10.96$\pm$0.05 &  7.81$\pm$0.05 \\
 10.0-10.5 &  187(61\%) &  10.67$\pm$0.07 &  7.50$\pm$0.07 &    1263(57\%) &  10.86$\pm$0.03 &   7.61$\pm$0.03 &    820(68\%) &  10.94$\pm$0.03 &  7.94$\pm$0.03 \\
 10.5-11.0 &   68(75\%) &  11.10$\pm$0.07 &  7.97$\pm$0.08 &     476(71\%) &  10.93$\pm$0.03 &   7.93$\pm$0.03 &    282(76\%) &  11.05$\pm$0.03 &  8.12$\pm$0.03 \\
    
    \multicolumn{10}{l}{(c) LT-SFH}\\
    \midrule
       8.5-9.0 &   48(86\%) &  5.64$\pm$0.26 &  2.77$\pm$0.19 &      81(77\%) &   5.76$\pm$0.20 &   3.00$\pm$0.14 &     21(70\%) &  6.97$\pm$0.36 &  3.47$\pm$0.26 \\
   9.0-9.5 &  119(75\%) &  5.55$\pm$0.17 &  2.76$\pm$0.12 &     246(75\%) &   5.52$\pm$0.11 &   2.85$\pm$0.08 &     61(61\%) &  7.20$\pm$0.18 &  4.28$\pm$0.14 \\
  9.5-10.0 &  236(65\%) &  5.86$\pm$0.10 &  3.04$\pm$0.07 &     898(60\%) &   6.69$\pm$0.05 &   3.64$\pm$0.04 &    392(46\%) &  8.71$\pm$0.06 &  5.74$\pm$0.05 \\
 10.0-10.5 &  122(39\%) &  7.92$\pm$0.10 &  5.19$\pm$0.08 &     936(43\%) &   8.28$\pm$0.04 &   5.34$\pm$0.03 &    392(32\%) &  9.40$\pm$0.04 &  6.64$\pm$0.04 \\
 10.5-11.0 &   23(25\%) &  8.90$\pm$0.19 &  6.17$\pm$0.16 &     194(29\%) &   9.15$\pm$0.06 &   6.28$\pm$0.05 &     91(24\%) &  9.67$\pm$0.06 &  7.03$\pm$0.07 \\    
    \end{tabular}

\end{minipage}
\end{center}
\caption{{\bf Median assembly times.} Assembly time (in Giga years, Gyr) of 50\% ($T_{\rm 50}$) and 70\% ($T_{\rm 70}$) of the stellar mass for all the SFH types (a), galaxies with ST-SFH (b), and LT-SFH (c), in voids, filaments \& walls, and clusters. The used samples take into account the selection criteria applied by the quality control analysis and are based on the same stellar mass distributions inside every stellar mass bin. The error represent the s.e.m. ($1\sigma$). The number of galaxies (n) in each environment and stellar mass bin is shown together with the fraction (\%) of each SFH type inside every stellar mass bin and environment. These are the values represented in Figure \ref{fig:T_KS}.\label{tab:Tf_KS}}
\end{sidewaystable}

    


\begin{figure*}
    \centering
        \includegraphics[width=\linewidth]{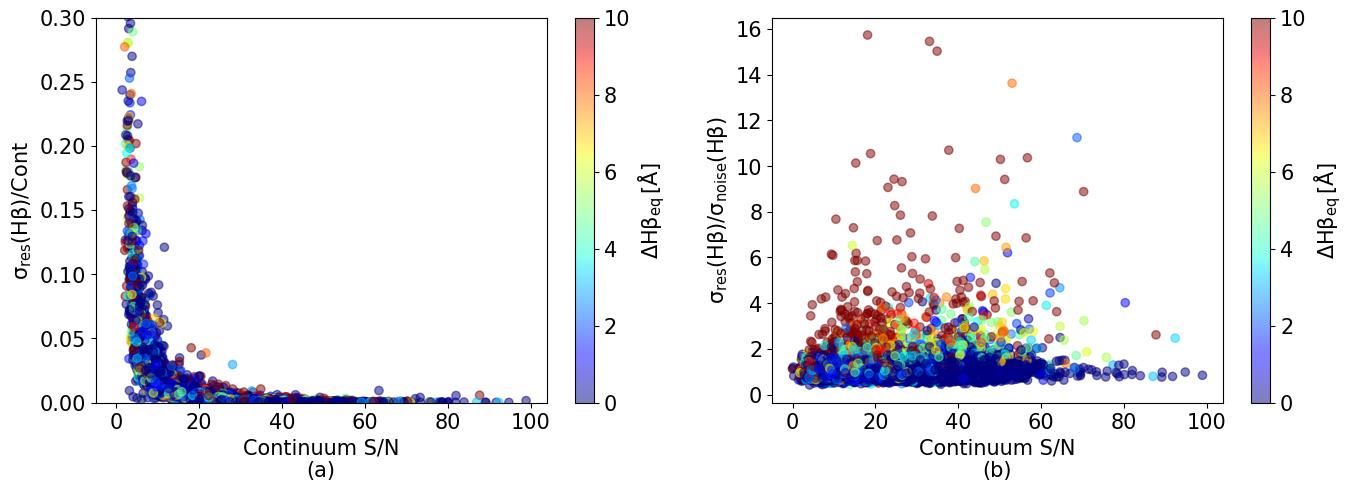}
    \caption{{\bf Fit residuals vs. spectral signal-to-noise, and emission lines.} (a) Standard deviation of the spectral fit residual ($\rm \sigma_{res}(H\beta)$) normalised by the level of the continuum (Cont) around $\rm H\beta$ vs. the spectral signal-to-noise (S/N) ratio in the continuum. (b) Residual-to-noise ratio as ($\rm \sigma_{res}(H\beta)$) normalised by standard deviation of noise in the continuum next to $\rm H\beta$ ($\rm \sigma_{noise}(H\beta)$) vs. the S/N in the continuum. The $\rm H\beta$ equivalent with ($\rm \Delta H\beta_{eq}$) is colour-coded in both panels.}
    \label{fig:sn_res}
    \end{figure*}

\begin{figure*}
    \centering
        \includegraphics[width=\linewidth]{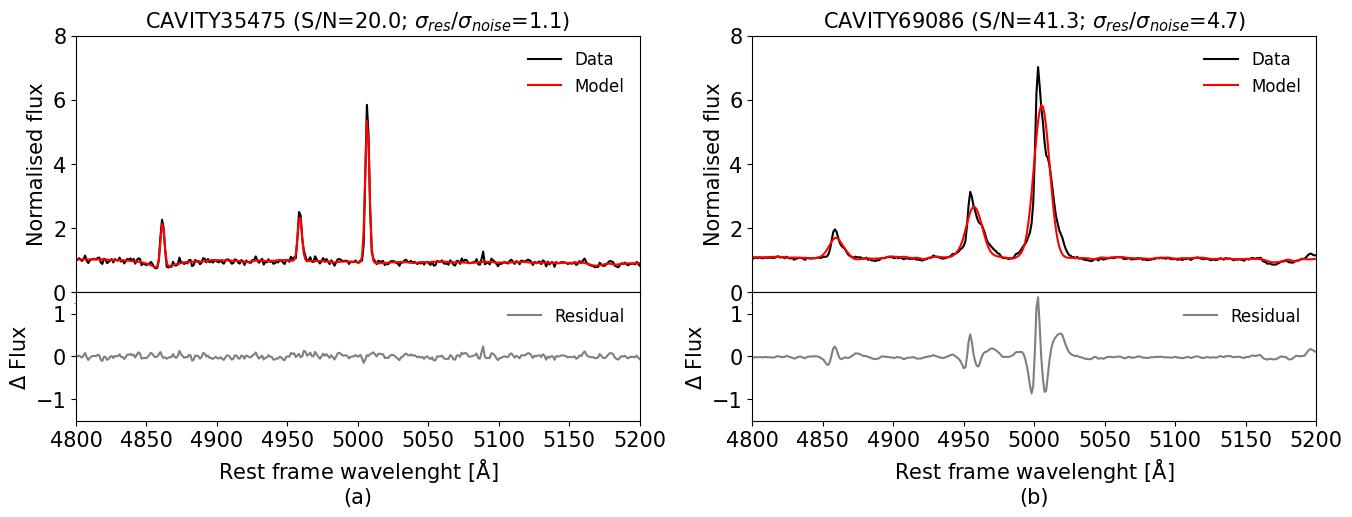}
    \caption{{\bf Examples of pPXF spectral fit of emission lines.} (a) Good fit example of a galaxy with signal-to-noise ratio of $\rm S/N =20.0$ and residual-to-noise ratio of $\rm \sigma_{res}/\sigma_{noise}=1.1$. (b) Bad fit example of a galaxy with $\rm S/N  =41.3$ and $\rm \sigma_{res}/\sigma_{noise}=4.7$. The black and red lines represent the observed and the fitted spectrum of the galaxy, respectively. The grey lines represent the fit residuals.}
    \label{fig:view_fit}
    \end{figure*}

\begin{figure}
    \centering
        \includegraphics[width=\linewidth]{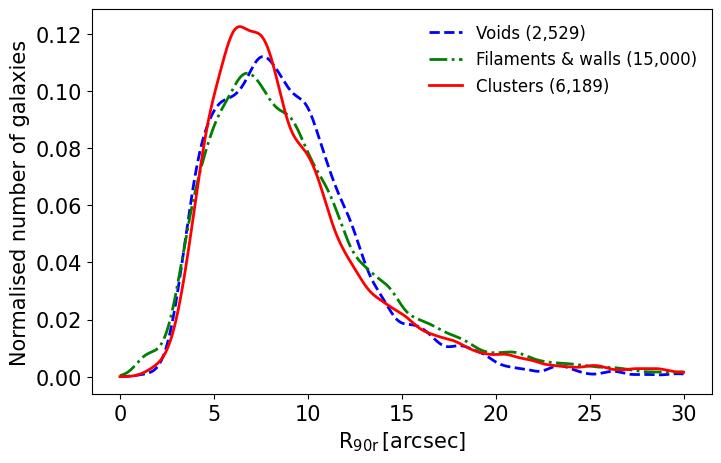}
    \caption{{\bf Distribution of apparent radius.} Normalised number of galaxies as a function of the apparent radius ($\rm R_{90r}$ from SDSS) for galaxies in voids (blue dashed line), filaments \& walls (green dot-dashed line), and clusters (red solid line) before the quality control. The apparent radius of the galaxies is represented by the petrosian radius containing the 90\% of the total flux of the galaxy in  r band (SDSS\cite{2020ApJS..249....3A}).}
    \label{fig:hist_R90}
    \end{figure}

\begin{figure*}
    \centering
        \includegraphics[width=\linewidth]{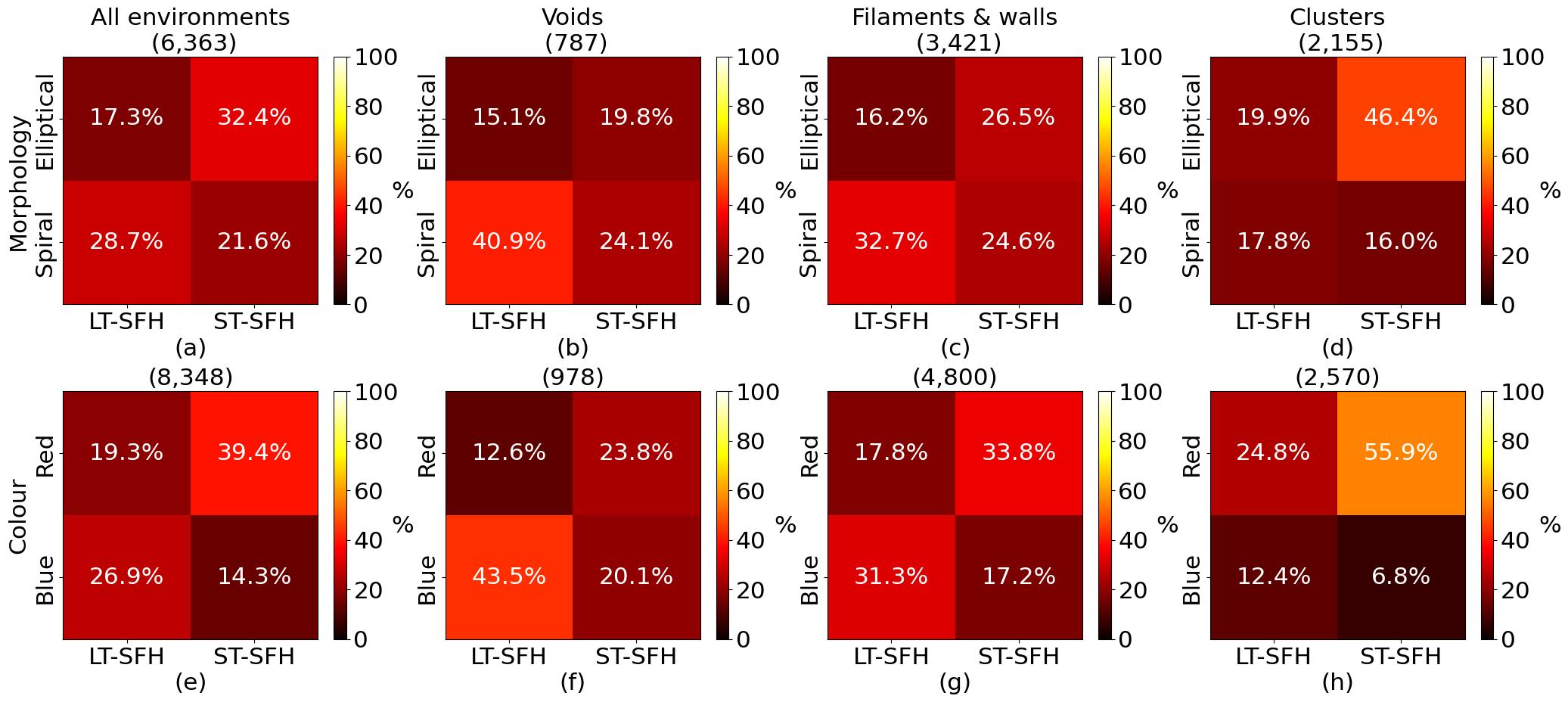}
    \caption{{\bf Correlation between the star formation history type, current morphology and colour of the galaxy.} Fraction of spiral and elliptical galaxies, or blue and red galaxies with LT-SFH and ST-SFH types is shown for all the environments together (a, e), for voids (b, f), filaments \& walls (c, g), and clusters (d, h), with the same stellar mass distribution. The number of galaxies is shown between brackets over each panel. Galaxies are blue if their $g-r<0.7$, red if $g-r>0.7$, spiral if T-type$>0$\cite{2018MNRAS.476.3661D}, and elliptical if T-type$<0$. Galaxies with ST-SFH are more likely to be elliptical or red. On the contrary, galaxies with LT-SFH are more likely to be spiral or blue. However, there is a significant fraction galaxies with ST-SFHs that are blue or spiral, and galaxies with LT-SFHs that are red or elliptical.}
    \label{fig:colour_mstar_types}
\end{figure*}

\begin{appendices}



\end{appendices}



\end{document}